\documentclass[10pt,a4paper,onecolumn]{article}
\usepackage{marginnote}
\usepackage{graphicx}
\usepackage{xcolor}
\usepackage{authblk,etoolbox}
\usepackage{titlesec}
\usepackage{calc}
\usepackage{tikz}
\usepackage{hyperref}
\hypersetup{colorlinks,breaklinks=true,
            urlcolor=[rgb]{0.0, 0.5, 1.0},
            linkcolor=[rgb]{0.0, 0.5, 1.0}}
\usepackage{caption}
\usepackage{tcolorbox}
\usepackage{amssymb,amsmath}
\usepackage{ifxetex,ifluatex}
\usepackage{seqsplit}
\usepackage{xstring}

\usepackage{float}
\let\origfigure\figure
\let\endorigfigure\endfigure
\renewenvironment{figure}[1][2] {
    \expandafter\origfigure\expandafter[H]
} {
    \endorigfigure
}

\usepackage{fixltx2e} 
\usepackage[
  backend=biber,
]{biblatex}
\bibliography{paper.bib}

\let\textttOrig=\texttt
\def\texttt#1{\expandafter\textttOrig{\seqsplit{#1}}}
\renewcommand{\seqinsert}{\ifmmode
  \allowbreak
  \else\penalty6000\hspace{0pt plus 0.02em}\fi}

\makeatletter
\let\href@Orig=\href
\def\href@Urllike#1#2{\href@Orig{#1}{\begingroup
    \def\Url@String{#2}\Url@FormatString
    \endgroup}}
\def\href@Notdoi#1#2{\def\tempa{#1}\def\tempb{#2}%
  \ifx\tempa\tempb\relax\href@Urllike{#1}{#2}\else
  \href@Orig{#1}{#2}\fi}
\def\href#1#2{%
  \IfBeginWith{#1}{https://doi.org}%
  {\href@Urllike{#1}{#2}}{\href@Notdoi{#1}{#2}}}
\makeatother

\usepackage[top=3.5cm, bottom=3cm, right=1.5cm, left=1.0cm,
            headheight=2.2cm, reversemp, includemp, marginparwidth=4.5cm]{geometry}

\titleformat{\section}
  {\normalfont\sffamily\Large\bfseries}
  {}{0pt}{}
\titleformat{\subsection}
  {\normalfont\sffamily\large\bfseries}
  {}{0pt}{}
\titleformat{\subsubsection}
  {\normalfont\sffamily\bfseries}
  {}{0pt}{}
\titleformat*{\paragraph}
  {\sffamily\normalsize}

\usepackage{fancyhdr}
\makeatletter
\let\ps@plain\ps@fancy
\fancyheadoffset[L]{4.5cm}
\fancyfootoffset[L]{4.5cm}

\definecolor{linky}{rgb}{0.0, 0.5, 1.0}

\newtcolorbox{repobox}
   {colback=red, colframe=red!75!black,
     boxrule=0.5pt, arc=2pt, left=6pt, right=6pt, top=3pt, bottom=3pt}

\newcommand{\ExternalLink}{%
   \tikz[x=1.2ex, y=1.2ex, baseline=-0.05ex]{%
       \begin{scope}[x=1ex, y=1ex]
           \clip (-0.1,-0.1)
               --++ (-0, 1.2)
               --++ (0.6, 0)
               --++ (0, -0.6)
               --++ (0.6, 0)
               --++ (0, -1);
           \path[draw,
               line width = 0.5,
               rounded corners=0.5]
               (0,0) rectangle (1,1);
       \end{scope}
       \path[draw, line width = 0.5] (0.5, 0.5)
           -- (1, 1);
       \path[draw, line width = 0.5] (0.6, 1)
           -- (1, 1) -- (1, 0.6);
       }
   }

\patchcmd{\@maketitle}{center}{flushleft}{}{}
\patchcmd{\@maketitle}{center}{flushleft}{}{}
\patchcmd{\@maketitle}{\LARGE}{\LARGE\sffamily}{}{}
\def\maketitle{{%
  
  \AB@maketitle}}
\makeatletter
\renewcommand\AB@affilsepx{ \protect\Affilfont}
\renewcommand\AB@affilnote[1]{{\bfseries #1}\hspace{3pt}}
\renewcommand{\affil}[2][]%
   {\newaffiltrue\let\AB@blk@and\AB@pand
      \if\relax#1\relax\def\AB@note{\AB@thenote}\else\def\AB@note{#1}%
        \setcounter{Maxaffil}{0}\fi
        \begingroup
        \let\href=\href@Orig
        \let\texttt=\textttOrig
        \let\protect\@unexpandable@protect
        \def\thanks{\protect\thanks}\def\footnote{\protect\footnote}%
        \@temptokena=\expandafter{\AB@authors}%
        {\def\\{\protect\\\protect\Affilfont}\xdef\AB@temp{#2}}%
         \xdef\AB@authors{\the\@temptokena\AB@las\AB@au@str
         \protect\\[\affilsep]\protect\Affilfont\AB@temp}%
         \gdef\AB@las{}\gdef\AB@au@str{}%
        {\def\\{, \ignorespaces}\xdef\AB@temp{#2}}%
        \@temptokena=\expandafter{\AB@affillist}%
        \xdef\AB@affillist{\the\@temptokena \AB@affilsep
          \AB@affilnote{\AB@note}\protect\Affilfont\AB@temp}%
      \endgroup
       \let\AB@affilsep\AB@affilsepx
}
\makeatother

\renewcommand\Affilfont{\sffamily\small\mdseries}
\ifnum 0\ifxetex 1\fi\ifluatex 1\fi=0 
  \usepackage[T1]{fontenc}
  \usepackage[utf8]{inputenc}

\else 
  \ifxetex
    \usepackage{mathspec}
    \usepackage{fontspec}

  \else
    \usepackage{fontspec}
  \fi
  \defaultfontfeatures{Ligatures=TeX,Scale=MatchLowercase}

\fi
\IfFileExists{upquote.sty}{\usepackage{upquote}}{}
\IfFileExists{microtype.sty}{%
\usepackage{microtype}
\UseMicrotypeSet[protrusion]{basicmath} 
}{}

\usepackage{hyperref}
\hypersetup{unicode=true,
            pdftitle={emcee v3: A Python ensemble sampling toolkit for affine-invariant MCMC},
            pdfborder={0 0 0},
            breaklinks=true}
\let\addcontentslineOrig=\addcontentsline
\def\addcontentsline#1#2#3{\bgroup
  \let\texttt=\textttOrig\addcontentslineOrig{#1}{#2}{#3}\egroup}
\let\markbothOrig\markboth
\def\markboth#1#2{\bgroup
  \let\texttt=\textttOrig\markbothOrig{#1}{#2}\egroup}
\let\markrightOrig\markright
\def\markright#1{\bgroup
  \let\texttt=\textttOrig\markrightOrig{#1}\egroup}

\usepackage{graphicx,grffile}
\makeatletter
\def\maxwidth{\ifdim\Gin@nat@width>\linewidth\linewidth\else\Gin@nat@width\fi}
\def\maxheight{\ifdim\Gin@nat@height>\textheight\textheight\else\Gin@nat@height\fi}
\makeatother
\setkeys{Gin}{width=\maxwidth,height=\maxheight,keepaspectratio}
\IfFileExists{parskip.sty}{%
\usepackage{parskip}
}{
\setlength{\parindent}{0pt}
\setlength{\parskip}{6pt plus 2pt minus 1pt}
}
\providecommand{\tightlist}{%
  \setlength{\itemsep}{0pt}\setlength{\parskip}{0pt}}
\ifx\paragraph\undefined\else
\let\oldparagraph\paragraph
\renewcommand{\paragraph}[1]{\oldparagraph{#1}\mbox{}}
\fi
\ifx\subparagraph\undefined\else
\let\oldsubparagraph\subparagraph
\renewcommand{\subparagraph}[1]{\oldsubparagraph{#1}\mbox{}}
\fi

\title{emcee v3: A Python ensemble sampling toolkit for affine-invariant MCMC}

        \author[1]{Daniel Foreman-Mackey}
          \author[1, 2]{Will M. Farr}
          \author[3, 4]{Manodeep Sinha}
          \author[5]{Anne M. Archibald}
          \author[1, 6]{David W. Hogg}
          \author[7]{Jeremy S. Sanders}
          \author[8]{Joe Zuntz}
          \author[9, 10]{Peter K. G. Williams}
          \author[11]{Andrew R. J. Nelson}
          \author[12]{Miguel de Val-Borro}
          \author[13]{Tobias Erhardt}
          \author[14]{Ilya Pashchenko}
          \author[15]{Oriol Abril Pla}
    
      \affil[1]{Center for Computational Astrophysics, Flatiron Institute}
      \affil[2]{Department of Physics and Astronomy, Stony Brook University, United
States}
      \affil[3]{Centre for Astrophysics \& Supercomputing, Swinburne University of
Technology, Australia}
      \affil[4]{ARC Centre of Excellence for All Sky Astrophysics in 3 Dimensions (ASTRO
3D)}
      \affil[5]{University of Newcastle}
      \affil[6]{Center for Cosmology and Particle Physics, Department of Physics, New
York University}
      \affil[7]{Max Planck Institute for Extraterrestrial Physics}
      \affil[8]{Institute for Astronomy, University of Edinburgh, Edinburgh, EH9 3HJ, UK}
      \affil[9]{Center for Astrophysics \textbar{} Harvard \& Smithsonian}
      \affil[10]{American Astronomical Society}
      \affil[11]{Australian Nuclear Science and Technology Organisation, NSW, Australia}
      \affil[12]{Planetary Science Institute, 1700 East Fort Lowell Rd., Suite 106,
Tucson, AZ 85719, USA}
      \affil[13]{Climate and Environmental Physics and Oeschger Center for Climate Change
Research, University of Bern, Bern, Switzerland}
      \affil[14]{P.N. Lebedev Physical Institute of the Russian Academy of Sciences,
Moscow, Russia}
      \affil[15]{Universitat Pompeu Fabra, Barcelona}
  \date{\vspace{-7ex}}

\begin{document}
\maketitle

\marginpar{

  \begin{flushleft}
  \sffamily\small

  {\bfseries DOI:} \href{https://doi.org/10.21105/joss.01864}{\color{linky}{10.21105/joss.01864}}

  \vspace{2mm}

  {\bfseries Software}
  \begin{itemize}
    \setlength\itemsep{0em}
    \item \href{https://github.com/openjournals/joss-reviews/issues/1864}{\color{linky}{Review}} \ExternalLink
    \item \href{https://github.com/dfm/emcee}{\color{linky}{Repository}} \ExternalLink
    \item \href{https://doi.org/10.5281/zenodo.3543502}{\color{linky}{Archive}} \ExternalLink
  \end{itemize}

  \vspace{2mm}

  \par\noindent\hrulefill\par

  \vspace{2mm}

  {\bfseries Editor:} \href{http://juanjobazan.com}{Juanjo Bazán} \ExternalLink \\
  \vspace{1mm}
    {\bfseries Reviewers:}
  \begin{itemize}
  \setlength\itemsep{0em}
    \item \href{https://github.com/benjaminrose}{@benjaminrose}
    \item \href{https://github.com/mattpitkin}{@mattpitkin}
    \end{itemize}
    \vspace{2mm}

  {\bfseries Submitted:} 28 October 2019\\
  {\bfseries Published:} 17 November 2019

  \vspace{2mm}
  {\bfseries License}\\
  Authors of papers retain copyright and release the work under a Creative Commons Attribution 4.0 International License (\href{http://creativecommons.org/licenses/by/4.0/}{\color{linky}{CC-BY}}).

  \end{flushleft}
}

\hypertarget{summary}{%
\section{Summary}\label{summary}}

\texttt{emcee} is a Python library implementing a class of
affine-invariant ensemble samplers for Markov chain Monte Carlo (MCMC).
This package has been widely applied to probabilistic modeling problems
in astrophysics where it was originally published (Foreman-Mackey, Hogg,
Lang, \& Goodman, 2013), with some applications in other fields. When it
was first released in 2012, the interface implemented in \texttt{emcee}
was fundamentally different from the MCMC libraries that were popular at
the time, such as \texttt{PyMC}, because it was specifically designed to
work with ``black box'' models instead of structured graphical models.
This has been a popular interface for applications in astrophysics
because it is often non-trivial to implement realistic physics within
the modeling frameworks required by other libraries. Since
\texttt{emcee}'s release, other libraries have been developed with
similar interfaces, such as \texttt{dynesty} (Speagle, 2019). The
version 3.0 release of \texttt{emcee} is the first major release of the
library in about 6 years and it includes a full re-write of the
computational backend, several commonly requested features, and a set of
new ``move'' implementations.

This new release includes both small quality of life improvements---like
a progress bar using \href{https://tqdm.github.io}{\texttt{tqdm}}---and
larger features. For example, the new \texttt{backends} interface
implements real time serialization of sampling results. By default
\texttt{emcee} saves its results in memory (as in the original
implementation), but it now also includes a \texttt{HDFBackend} class
that serializes the chain to disk using
\href{https://www.h5py.org}{h5py}.

The most important new feature included in the version 3.0 release of
\texttt{emcee} is the new \texttt{moves} interface. Originally,
\texttt{emcee} implemented the affine-invariant ``stretch move''
proposed by Goodman \& Weare (2010), but there are other ensemble
proposals that can get better performance for certain applications.
\texttt{emcee} now includes implementations of several other ensemble
moves and an interface for defining custom proposals. The implemented
moves include:

\begin{itemize}
\tightlist
\item
  The ``stretch move'' proposed by Goodman \& Weare (2010),
\item
  The ``differential evolution'' and ``differential evolution snooker
  update'' moves (ter Braak, 2006; ter Braak \& Vrugt, 2008), and
\item
  A ``kernel density proposal'' based on the implementation in
  \href{https://github.com/bfarr/kombine}{the \texttt{kombine} library}
  (Farr \& Farr, 2015).
\end{itemize}

\texttt{emcee} has been widely used and the original paper has been
highly cited, but there have been many contributions from members of the
community. This paper is meant to highlight these contributions and
provide citation credit to the academic contributors. A full up-to-date
list of contributors can always be found
\href{https://github.com/dfm/emcee/graphs/contributors}{on GitHub}.

\hypertarget{references}{%
\section*{References}\label{references}}
\addcontentsline{toc}{section}{References}

\hypertarget{refs}{}
\leavevmode\hypertarget{ref-Farr:2015}{}%
Farr, B., \& Farr, W. M. (2015). Kombine: A kernel-density-based,
embarrassingly parallel ensemble sampler. Retrieved from
\url{https://github.com/bfarr/kombine}

\leavevmode\hypertarget{ref-Foreman-Mackey:2013}{}%
Foreman-Mackey, D., Hogg, D. W., Lang, D., \& Goodman, J. (2013). emcee:
The MCMC Hammer. \emph{Publications of the Astronomical Society of the
Pacific}, \emph{125}(925), 306.
doi:\href{https://doi.org/10.1086/670067}{10.1086/670067}

\leavevmode\hypertarget{ref-Goodman:2010}{}%
Goodman, J., \& Weare, J. (2010). Ensemble samplers with affine
invariance. \emph{Communications in applied mathematics and
computational science}, \emph{5}(1), 65--80.
doi:\href{https://doi.org/10.2140/camcos.2010.5.65}{10.2140/camcos.2010.5.65}

\leavevmode\hypertarget{ref-Speagle:2019}{}%
Speagle, J. S. (2019). dynesty: A Dynamic Nested Sampling Package for
Estimating Bayesian Posteriors and Evidences. \emph{arXiv e-prints},
arXiv:1904.02180. Retrieved from \url{http://arxiv.org/abs/1904.02180}

\leavevmode\hypertarget{ref-Ter-Braak:2006}{}%
ter Braak, C. J. F. (2006). A Markov Chain Monte Carlo version of the
genetic algorithm Differential Evolution: easy Bayesian computing for
real parameter spaces. \emph{Statistics and Computing}, \emph{16}(3),
239--249.
doi:\href{https://doi.org/10.1007/s11222-006-8769-1}{10.1007/s11222-006-8769-1}

\leavevmode\hypertarget{ref-Ter-Braak:2008}{}%
ter Braak, C. J. F., \& Vrugt, J. A. (2008). Differential evolution
Markov chain with snooker updater and fewer chains. \emph{Statistics and
Computing}, \emph{18}(4), 435--446.
doi:\href{https://doi.org/10.1007/s11222-008-9104-9}{10.1007/s11222-008-9104-9}

\end{document}